\documentclass[12pt]{iopart}

\usepackage{iopams}
\usepackage{graphicx}

\newcommand{\eqref}[1]{\eref{#1}}
 \newcommand{\bm}[1]{\bi{#1}}

  \def\dfrac{\displaystyle\frac}

\renewcommand {\Re}{\mathop{\mathrm{Re}}\nolimits}
\renewcommand {\Im}{\mathop{\mathrm{Im}}\nolimits}

\renewcommand {\phi}{\varphi}

\newcommand{\eps}{\varepsilon}

\begin{document}
  \title{
Purcell effect in wire metamaterials}

  \author{Alexander N. Poddubny,$^{1,2}$ Pavel A. Belov,$^{1,3}$ \\
 and Yuri S. Kivshar$^{1,4}$}

 \address{$^{1}$National Research University for Information Technology, Mechanics and Optics
(ITMO), St.~Petersburg 197101, Russia\\
$^{2}$Ioffe Physical-Technical Institute of the Russian Academy of Sciences,
St.~Petersburg 194021, Russia\\
$^{3}$School of Electronic Engineering and Computer Science, Queen Mary University of London, London E1 4NS, UK\\
$^{4}$Nonlinear Physics Center and Center for Ultrahigh-bandwidth Devices for
Optical Systems (CUDOS), Research School of Physics and Engineering, Australian National University,
Canberra ACT 0200, Australia}
\ead{a.poddubny@phoi.ifmo.ru}
\begin{abstract}
We study theoretically the enhancement of spontaneous emission in wire metamaterials.
We analyze the dependence of the Purcell factor dependence on wire dielectric constant for 
both electric and magnetic dipole sources, and find an optimal value of the dielectric constant 
for maximizing the Purcell factor for the electric dipole. We obtain analytical expressions for 
the Purcell factor and also provide estimates for the Purcell factor in realistic structures 
operating in both microwave and optical spectral range.
\end{abstract}

\maketitle

\section{Introduction}

Wire metamaterials are composed of arrays of optically thin metallic rods embedded in a dielectric matrix~\cite{simovski2012}. Experimental realizations of such structures span from microwaves \cite{belov2008trans} to optics~\cite{Wurtz2008, pollard2009,kabashin2009,narimanov2009b,Noginov2010}, and
they are very promising for a number of applications including the subwalength transmission of images~\cite{belov2008trans}, negative refraction phenomena~\cite{Yao2008}, superlensing~\cite{Fink2012},
and biosensing applications~\cite{kabashin2009}. Strong enhancement of the Vavilov-Cherenkov radiation and peculiar dipole emission patterns have been also predicted for wire metamaterials~\cite{vorobev2012,fernandes2012,maslovski2012}.

Specific property of wire metamaterials is the strong spatial dispersion of the effective dielectric constant~\cite{belov2003}, manifested as the formation of so-called TEM modes in addition to the TE and TM modes of an ordinary uniaxial structure. Due to a finite value of the dielectric constant of wires, TEM modes acquire hyperbolic dispersion~\cite{silveirinha2006}. In this regard the wire medium represents a particular class of metamaterials with hyperbolic isofrequency surfaces, which have recently attracted a lot of attention~\cite{shalaev2011,Krishnamoorthy2012}. The specific feature of the hyperbolic metamaterials is diverging photonic density of states, promoting high spontaneous decay rate of the embedded light source~\cite{poddubny2011pra,sipe2011}. In wire metamaterials this effect should lead to large Purcell factor as well. 

Although the  Purcell factor in hyperbolic metamaterials stays finite due to the discreteness of the actual structure~\cite{Iorsh2012,kidwai2012,poddubny2012cross,cortes2012}, its value is strongly sensitive to the structure geometry. Consequently, the detailed theory recently developed for planar metal-dielectric metamaterials \cite{sipe2011,Iorsh2012,kidwai2012} is not applicable directly to nanowire arrays and novel study is required. The crude estimation of the Purcell factor is provided by the
density of states enhancement of TEM modes$\sim(\lambda/a)^2$~\cite{maslovski2011}, where $\lambda$ is the wavelength and $a$ is the structure period. Our goal is to perform more comprehensive analysis which takes into account finite dielectric constant of the wires and arbitrary spatial position of the source in the metamaterial unit cell. We  consider  both electric and magnetic dipole sources, because high sensitivity of the enhancement factor to the dipole source is known already since the original work of Purcell~\cite{Purcell,Glazov2011}.

\begin{figure}[b]
\centering\includegraphics[width=0.6\linewidth]{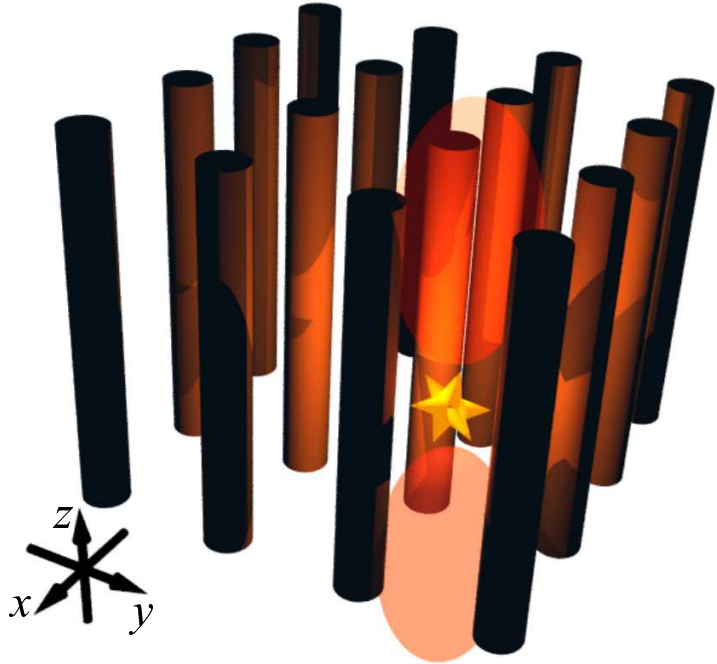}
 \caption{Schematic illustration of the wire metamaterial with embedded light source.
}\label{fig:wires}
\end{figure}

The paper is organized as follows. Section~\ref{sec:model} presents our theoretical model. Section~\ref{sec:disp} is devoted to the analysis of the eigenmode dispersion of the structure. Section~\ref{sec:purc} contains numerical and analytical results for the Purcell factor. Discussion of the Purcell factor attainable in different experimental conditions is summarized in Sec.~\ref{sec:Zayats}.
whereas the last section concludes the paper. 

\section{Local field approach}\label{sec:model}

The structure under consideration is  illustrated schematically in Fig.~\ref{fig:wires}. It consists of identical infinitely long wires with radius $R$, arranged in infinite square lattice with period $a$ and parallel to $z$ axis. Wires are characterized with dielectric constant $\varepsilon_{\rm wire}$. Electric ($\bm p$) or magnetic ($\bm m$) dipole  source  is embedded in the structure at the point $\bm r_0\equiv\boldsymbol \rho_0 +z_0\hat {\bi z}$. Electric field  satisfies the following equation:
\begin{equation}\label{eq:E}
 \nabla\times\nabla\times\bm E-q^2\varepsilon(\boldsymbol \rho)\bm E=4\pi q^2\bm P
\end{equation}
where
\begin{equation}\label{eq:epsilon}
 \varepsilon(\boldsymbol \rho)=1+(\varepsilon_{\rm wire}-1)\sum\limits_{j}\theta(R-|\boldsymbol\rho-\boldsymbol \rho_j|)
\end{equation}
is the dielectric function (here $\theta$ is the Heaviside step function),
\begin{equation}
 \bm P=\cases{
        \bm p\:\delta(\bm r-\bm r_0),\\
        \frac{\rmi}{q}\nabla\times\bm m\delta(\bm r-\bm r_0).}
\end{equation}
is the polarization term, describing the point electric or magnetic dipole source, and $q=\omega/c$.
We neglect the transverse polarizability of the wires assuming that they are thin enough. Therefore, wire polarization per unit length is solely described by axial polarizability $\alpha$, determined from
\begin{equation}
 \bm P(z)=\int \frac{\rmd k_z}{2\pi}\e^{\rmi k_zz}\alpha(k_z)\hat{\bm z}E_z(k_z)\:,
\end{equation}
and given by\cite{belov2002b,belov2003,silveirinha2006}
\begin{equation}\label{eq:alpha}
 \frac{1}{\alpha}\approx -\rmi \pi q_\perp^2 H_0^{(1)}\left(q_\perp R\right)+\frac{4}{(\varepsilon_{\rm wire}-1)R^2}\:,
\end{equation}
where $q_{\perp}^2=q^2-k_z^2$.  Essential feature of Eq.~\eqref{eq:alpha} is the  spatial dispersion of the wire dielectric responce, i.e., the dependence of the polarizability on the wavevector $k_z$.

Purcell factor for electric and magnetic dipole emission may be found via the imaginary part of the field, induced in the structure by the source\cite{Ivchenko2005,Novotny2006,Glazov2011}
\begin{eqnarray}
 f^{(e)}&=\frac{3}{2q^3p^2}\Im [\bm E(\bm r_0)\cdot\bm p]\:,\label{eq:PurcEM}\\
f^{(m)}&=\frac{3}{2q^3m^2}\Im [\bm H(\bm r_0)\cdot\bm m]\nonumber\:.
\end{eqnarray}
The field is given by a sum over Bloch waves with wavevectors $\bm k=\bm k_{\perp}+k_z\hat{\bm z}$, similar as can be done  for a 3D dipole lattice~\cite{poddubny2012cross}. The answer reads
\begin{eqnarray}\label{eq:f}
 f=1+\frac{3 }{2q^3}\Im
\int \frac{S\rmd^3 k}{(2\pi)^3}
\frac{[\bm G_{\bm k}(\boldsymbol{\rho}_0)\cdot \bm n] [\bm G_{\bm k}(-\boldsymbol{\rho}_0)\cdot \bm n]}{1/\alpha(\bm k)-C(\bm k)-\rmi 0}\:.
\end{eqnarray}
Here $S=a^2$ is the unit cell area, the integration over
$k_x$ and $k_y$ is performed over the Brillouin zone from $-\pi/a$ to $\pi/a$, the integration range over wavevector $k_z$ extends from $-\infty$ to $\infty$
and $\bm n$ is the unit vector oriented along the dipole.
The quantity $C$ in \eqref{eq:f} is the interaction constant of the wires~\cite{belov2002b}, defined as
\begin{equation}\label{eq:C}
 C(\bm k)=\rmi\pi q_{\perp}^2\sum\limits_{\bm r_j\ne 0}\e^{\rmi \bm k_{\perp}\boldsymbol \rho_j}H_0^{(1)}(q\rho_j)\:,
\end{equation}
and
 \begin{equation}\label{eq:Gk}
  \bm G_{\bm k}(\boldsymbol \rho)=\sum\limits_{j}\bm G_0(\boldsymbol \rho-\boldsymbol \rho_j,k_z)\e^{\rmi \bm k(\boldsymbol \rho-\boldsymbol \rho_j)}
 \end{equation}
is the periodic Green function for waves with Bloch vector $\bm k_{\perp}$.
For electric and magnetic dipole problems one should substitute in \eqref{eq:Gk}
for $\bm G_{0}$
electric and magnetic fields of single polarized wire
 $\bm G_0^{(e)}$ and $\bm G_0^{(m)}$, respectively. Corresponding fields are given by
\begin{eqnarray}
 \bm G_0^{(e)}(\boldsymbol \rho,k_z)=  \rmi \pi
 q_\perp^2 \hat{\bm z}H_0^{(1)}(q_{\perp}\rho)+\pi k_z q_\perp
 \hat{\boldsymbol\rho}
 H_1^{(1)}(q_{\perp}\rho)\label{eq:Gek}\:,
\end{eqnarray}
for electric dipole and
\begin{equation}
 \bm G_0^{(m)}(\boldsymbol \rho,k_z)= \pi q_{\perp} q\hat{\boldsymbol \phi}H_1^{(1)}(q_{\perp}\rho)\:,\label{eq:Gmk}
\end{equation}
for magnetic one. Eqs.~\eqref{eq:f}--\eqref{eq:Gmk} present a general explicit answer for Purcell factors. In practical calculations, however, the series \eqref{eq:C},\eqref{eq:Gk} should be evaluated not directly but using either Ewald summation~\cite{Wang1993} or Floquet-type summation~\cite{belov2002b}. The latter approach turns out to be more numerically efficient and yields the following result \cite{belov2002b,belov2003}
\begin{eqnarray}
 C(\bm k)=-2q_\perp^2\ln\frac{2\pi R}{a}\nonumber\\-\sum\limits_{m=-\infty}^\infty\left(\frac{2\pi q_{\perp}^2}{q_{x,m}a}\frac{\sin q_{x,m}a}{
\cos q_{x,m}a-\cos k_x a
}-\frac{q_{\perp}^2(1-\delta_{m,0})}{|m|}\right)\:\label{eq:C_Floquet}
\end{eqnarray}
for the interaction constant, where
$q_{x,m}=\sqrt{q^2-k_z^2-k_{y,m}^2}$ and $k_{y,m}=k_{y}+2\pi m/a$.
Similar technique can be used to calculate the Green function, result reads
\begin{equation}
\bm G^{(e)}_{\bm k}(\boldsymbol \rho)=\sum\limits_m
  [-q_{\perp}^2\mathcal S_m\hat{\bm z}- \rmi  k_zq_{x,m} \mathcal C_m \hat{\bm x}+
 k_{y,m}k_z\mathcal S_m\hat{\bm y}]\:. \label{eq:Gk_Floq}
\end{equation}
where
\begin{eqnarray}
\mathcal S_m=\frac{2\pi }{q_{x,m}a}\e^{\rmi k_{y,m}y}\frac{\e^{\rmi k_x a}\sin q_{x,m}x-\sin q_{x,m}(x-a)}{\cos k_xa-\cos q_{x,m}a},\label{eq:Cm}\\
\mathcal C_m=\frac{2\pi }{q_{x,m} a}\e^{\rmi k_{y,m}y}\frac{\e^{\rmi k_x a}\cos q_{x,m}x-\cos q_{x,m}(x-a)}{\cos k_xa-\cos q_{x,m}a}\:.\nonumber
\end{eqnarray}
The Green function $\bm G_{\bm k}^{(m)}$ may be obtained from Eqs.~\eqref{eq:Gk_Floq},\eqref{eq:Cm} using the following expression
\begin{equation}
 \bm G_{\bm k}^{(m)}=\frac1{q}(\hat{\bm x}\partial_x+\hat{\bm y}\partial_y+\rmi  k_z\hat{\bm z})\times
\bm G_{\bm k}^{(e)}\:.
\end{equation}

Below we present numerical results for Purcell factor  along with analytical answers in certain limiting cases along with. We first determine analytical expressions for the eigenmodes of the structure, Sec.~\ref{sec:disp}, and then analyze Purcell factor  dependence on the source position, Sec.~\ref{sec:close}, and on the dielectric constant of the wires, Sec.~\ref{sec:epsilon}.
\section{Dispersion analysis} \label{sec:disp}
\begin{figure}[t]
 \includegraphics[width=\linewidth]{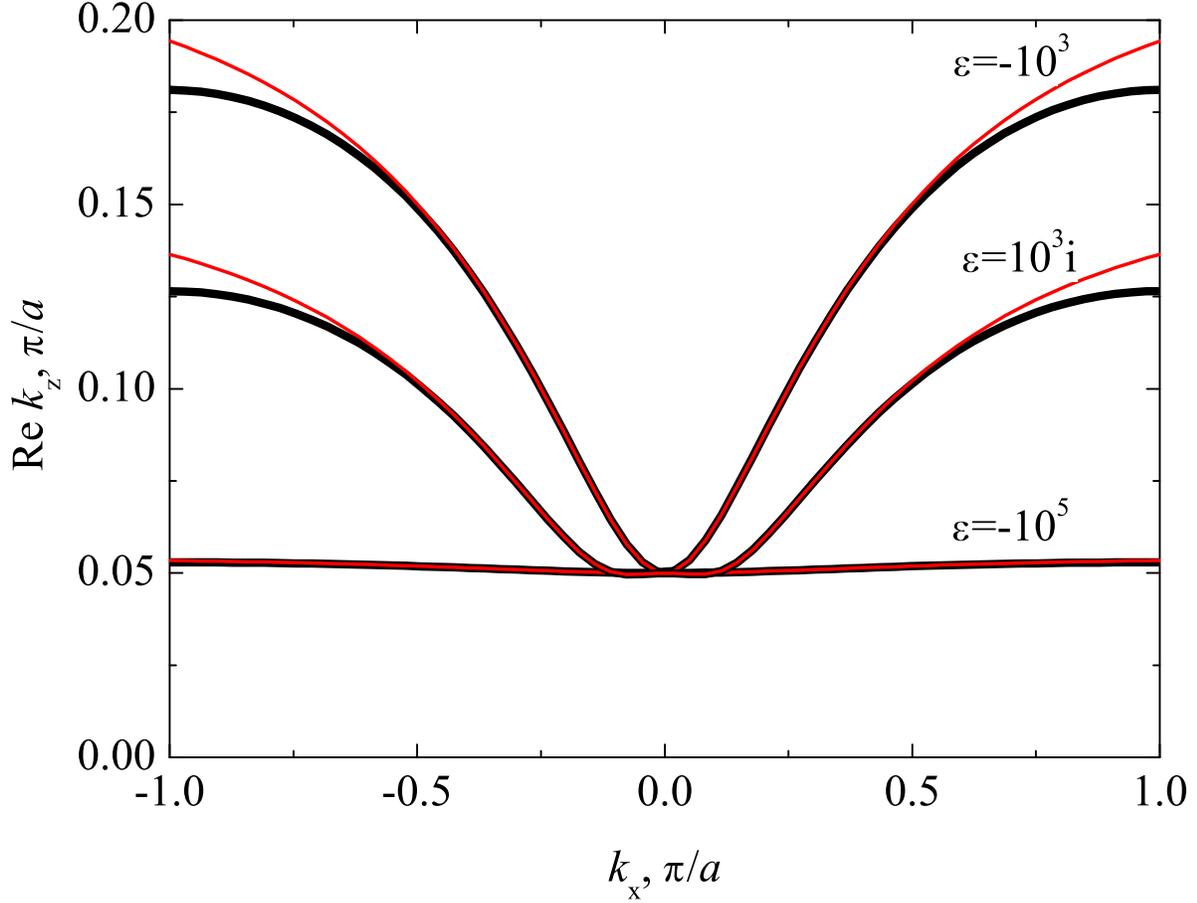}
\caption{Isofrequency curves for quasi-TEM modes calculated for the  different values of the wire dielectric constant $\varepsilon_{\rm wire}$. Thick curves are numerical results,  thin curves are plotted according to Eq.~\eqref{eq:kz12b}. Calculation was performed at $qa=0.05\pi$ and $R/a=0.05$.}\label{fig:kz}
\end{figure}
The spontaneous decay is due to the emission of photons, with energies found from the zeros of the resonant denominator in Eq.~\eqref{eq:f}:
$1/\alpha-C=0$.
Using Eqs.~\eqref{eq:alpha},\eqref{eq:C_Floquet}
we can present  this denominator for $qa\ll1$ and $ka\ll 1$
as
\begin{equation}\label{eq:nice2}
 \frac{1}{\alpha}-C\approx -\frac{4\pi }{a^2 q_p^2}\left[(q^2-k_z^2)\left(\frac{q_p^2}{k^2-q^2}+1\right)+\varkappa^2\right]\:,
\end{equation}
where
\begin{equation}
  \varkappa^2=q_p^2\frac{a^2}{\pi(1-\eps_{\rm wire}1)R^2},\quad
 \frac1{q_p^2}\approx
   \frac{a^2}{2\pi}\left[\ln\left(\dfrac{a}{2\pi R}\right)+\dfrac{\pi}{6}\right]\:.
\end{equation}
Here the wavevector $\varkappa\propto 1/\sqrt{\eps_{\rm wire}-1}$ characterizes the nonperfect character of the wires and $q_p$ is the effective plasma wavevector.
The physical sense of these parameters is better understood when Eq.~\eqref{eq:nice2} is rewritten as
\begin{equation}\label{eq:nice3}
 \frac{1}{\alpha}-C\approx- \frac{4\pi}{a^2 q_p^2}
\frac{(k_z^2-k_{1}^2)(k_z^2-k_{2}^2)}{k_{\rm TM}^2+q_p^2-k_z^2}\:,
\end{equation}
where
\begin{eqnarray}\label{eq:k12z}
 k_{1,2}^2=\frac{q^2+\varkappa^2+k_{\rm TM}^2}{2}\\\pm\sqrt{\frac{1}{4}
\left(q^2+\varkappa^2-k_{\rm TM}^2\right)^2
-q_p^2\varkappa^2}
\end{eqnarray}
are the wavevectors of the eigenmodes of the wire medium and
\begin{equation}\label{eq:kTEM}
 k_{\rm TM}=\sqrt{q^2-k_\perp^2-q_p^2}\:.
\end{equation}
Eigenmodes~\eqref{eq:k12z} can be also obtained if the wire medium is treated as homogeneous  medium with spatial dispersion\cite{silveirinha2006},
where effective dielectric constant reads
\begin{equation}\label{eq:eps_eff}
\varepsilon_{xx}=\eps_{yy}=1,\quad \eps_{zz}=1-\frac{q_p^2}{q^2-\varkappa^2-k_z^2}\:.
\end{equation}
For large values of
wire dielectric constant ($\varepsilon_{\rm wire}\gg 1$),
 when  $\varkappa$ is small, Eqs.~\eqref{eq:k12z} may be approximately rewritten as
\begin{equation}\label{eq:kz12b}
 k_{1}^2\approx q^2+\frac{\varkappa^2k_\perp^2}{q_p^2+k_\perp^2},\quad
k_{2}^2\approx q^2-q_p^2-k_\perp^2+\frac{\varkappa^2q_p^2}{q_p^2+k_\perp^2}\:.
\end{equation}
For perfect wires with $\eps_{\rm wire}\to \infty$ Eqs.~\eqref{eq:kz12b} reduce to
\begin{eqnarray}\label{eq:kTEM2}
& k_{1}=q,&\mbox{(TEM mode),}\\
&k_{2}=k_{\rm TM},&\mbox{(TM mode).}\nonumber
\end{eqnarray}

The dispersionless TEM modes are specific for wire medium~\cite{belov2003}. For $q<q_p$ they are the only propagating modes in the structure and thus solely control spontaneous emission rate.
Due to the high density of TEM photonic states one can also expect large Purcell factor~\cite{maslovski2011}.
For non-perfect wires TEM and TM modes mix, as can be seen from Eqs.~\eqref{eq:k12z}.

Fig.~\ref{fig:kz} presents isofrequency curves of quasi-TEM modes $k_{1}^{(z)}$ for different values of wire dielectric constant. For finite values of $\varepsilon_{\rm wire}$ the TEM modes acquire hyperbolic-like dispersion and the absolute values  of the wavevector $k_z$ increase according to Eq.~\eqref{eq:kz12b}.
The growth of the wavevector for non-perfect wires is illustrated on Fig.~\ref{fig:plasma}a.
Another effect of the finite dielectric constant is the decrease of the effective plasma wavevector, defined as the cutoff of TM waves
\begin{equation}\label{eq:plasma_eff}
 \tilde q^2_p\equiv q^2-k^2_{2}(k_{\perp}=0)\approx q_p^2-\varkappa^2\:.
\end{equation}
Eq.~\eqref{eq:plasma_eff} indicates, that the plasma frequency
$c\tilde q_p$ decreases for non-perfect wires. Corresponding dependence  is shown in Fig.~\ref{fig:plasma}b.  For  small enough values of $\varepsilon_{\rm wire}\sim (a/R)^2$ the plasma frequency turns to zero so the wire medium loses its properties.
\begin{figure}[t]
 \includegraphics[width=\linewidth]{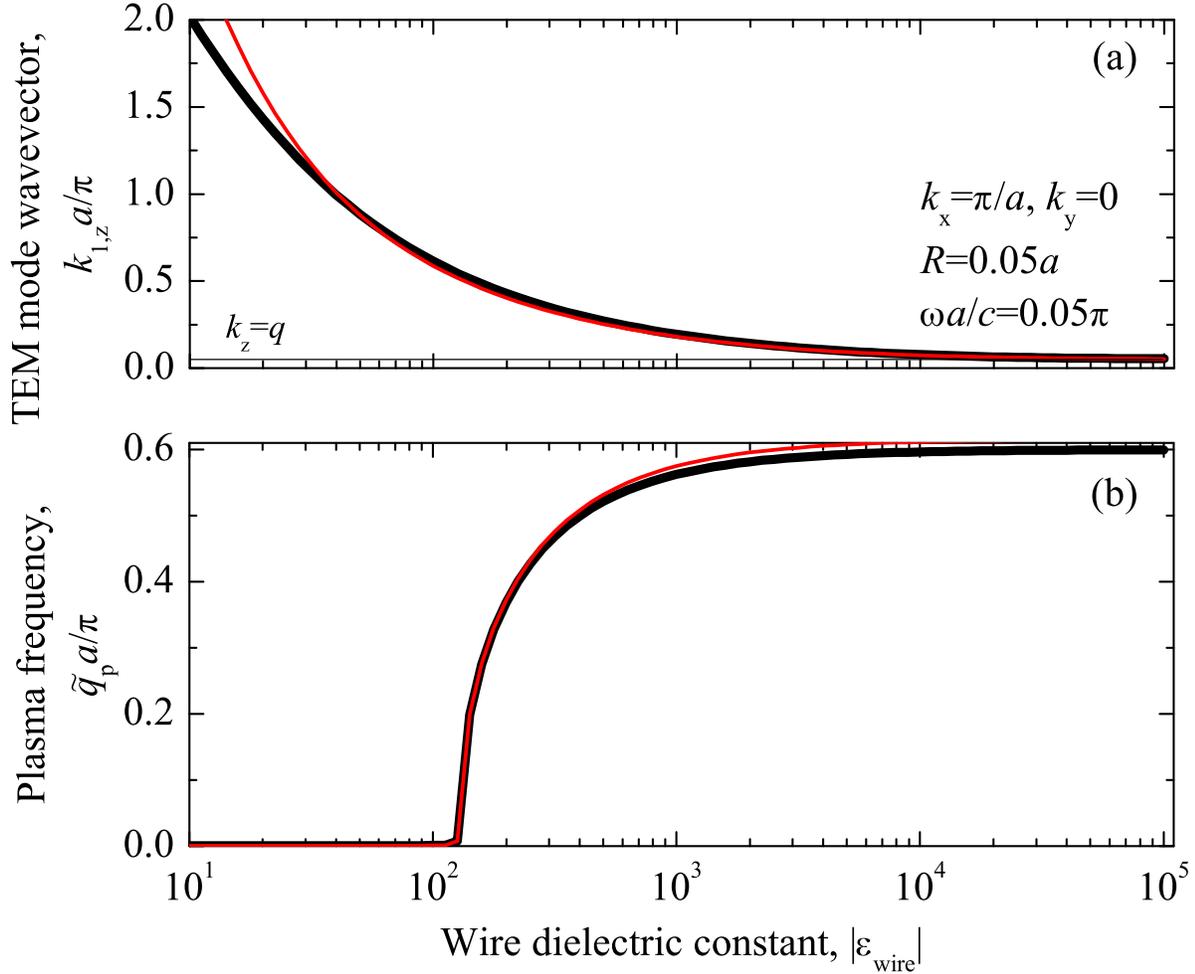}
\caption{(Color online) Effective plasma frequency $\tilde q_p$ (a) and quasi-TEM mode wavevector (b) as functions of the wire dielectric constant.
Thick black and thin red lines correspond to numerical calculation and Eqs.~\eqref{eq:kz12b}, respectively.
  Calculation was performed for $\eps_{\rm wire}=-|\eps_{\rm wire}|$, $qa=0.05\pi$, $R/a=0.05$, and $x_0=y_0=a/2$.}\label{fig:epsilon}
\end{figure}

\section{Purcell factor}\label{sec:purc}
Now we proceed to the dependence of the Purcell factor
 on the dipole position within the unit cell of the structure and
 on the wire dielectric constant.
\subsection{Effect of dipole position}\label{sec:close}
In this section we focus on perfect wires with
$\varepsilon_{\rm wire}\to\infty$
and analyze, how the Purcell factor depends on the coordinate $\boldsymbol \rho_0=(x_0,y_0)$ of the light source.
We will first obtain analytical results and then  compare them with numerical calculations.
For perfect wires the integral over $k_z$ in \eqref{eq:f} is determined by the residue at the wavevector corresponding to the TEM modes:
\begin{equation}\label{eq:intkz}
 \Im \int \frac{\rmd k_z}{2\pi} \frac{|G(k_z)|^2}{1/\alpha-C-\rmi0}=
\frac{|G(q)|^2}{2q}\frac{a^2}{4\pi}\dfrac1{\dfrac1{k_\perp^2}+\dfrac1{q_p^2}}\:.
\end{equation}
In Eq.~\eqref{eq:intkz} we have used the representation \eqref{eq:nice3} of the resonant denominator $1/\alpha-C$.
For dipole close enough to the wires the dipole the Bloch Green function
\eqref{eq:Gk}
is mainly determined by the field of the nearest wire
\begin{equation}
 \bm G_{\bm k}(\boldsymbol \rho)\approx \bm G_{0}(\boldsymbol \rho)\:.
\end{equation}
For electric and magnetic dipole this leads to
\begin{equation}\label{eq:kzTEM}
 \bm G^{(e)}(k_z=q)\approx -2\rmi \frac {q\hat{\boldsymbol \rho}}{\rho}\e^{\rmi qz},\quad
 \bm G^{(e)}(k_z=q)\approx -2\rmi \frac {q\hat{\boldsymbol \phi}}{\rho}\e^{\rmi qz}
\:.
\end{equation}
Eq.~\eqref{eq:kzTEM} indicates, that  for perfect wires only in-plane electric and magnetic dipoles couple with TEM waves. Thus, only for this orientation Purcell factor is enhanced due to the large density of TEM waves.
Substituting Eq.~\eqref{eq:kzTEM} and Eq.~\eqref{eq:intkz} into Eq.~\eqref{eq:f} we obtain
\begin{equation}\label{eq:fe1}
 f^{(e)}=\frac{3a^4q_p^2}{8\pi^2q^2}\frac{(\hat{\boldsymbol \rho}_0\cdot\bm n)^2}{\rho_0^2}\int\limits_0^{K_{\rm max}}\frac{k^3\rmd k}{k^2+q_p^2}\:.
\end{equation}
Here $K_{\rm max}$ is the in-plane wavevector cutoff, steaming from the finite size of the Brillouin zone. We use the value $K_{\max}=2\sqrt{\pi}/a$, so that the size of the Brillouin zone in cylindrical approximation stays the same, $\pi K_{\rm max}^2=(2\pi/a)^2$.
Performing the integration in Eq.~\eqref{eq:fe1} we obtain
\begin{eqnarray}\label{eq:f_e_pos}
 f^{(e)}=\frac{(\hat{\boldsymbol \rho}_0\cdot \bm n)^2a^2}{\rho_0^2}\frac{3}{16\pi^2q^2a^2}\\\times\left[K_{\rm max}^2q_p^2a^4-q_p^4a^4\ln\left(1+\frac{K_{\rm max}^2}{q_p^2}\right)\right]\:.
\end{eqnarray}
For magnetic dipole one has
\begin{equation}\label{eq:f_m_pos}
 f^{(m)}(\bm n)=f^{(e)}(\bm n\times\hat{\bm z})\:,
\end{equation}
i.e. Purcell factor for electric dipole, parallel to radius-vector $\boldsymbol \rho_0$,  is the same as for the in-plane magnetic dipole, perpendicular to $\boldsymbol \rho_0$.
We note, that Eq.~\eqref{eq:f_m_pos} is the general  relation between the Purcell factors of electric and magnetic dipole for perfect wires, holding at any distance from the wires.
Eqs.~\eqref{eq:f_e_pos},\eqref{eq:f_m_pos} present the analytical answer for in-plane electric and magnetic dipoles positioned close to the wires. For vertical dipoles one has
$f_{z}^{(e)}=0$ and $f_{z}^{(m)}=1$ at $\eps_{\rm wire}\to\infty$, independent of the dipole coordinates.
The general structure of Eq.~\eqref{eq:f_e_pos} is the same as for the 3D arrays of resonant dipoles~\cite{poddubny2012cross}. It consists of two factors, the first factors describes the local field enhancement. Second factor describes  collective effect: density  of states enhancement due to the TEM modes. It can be estimated by the order of magnitude as $\lambda^2/a^2$, where $\lambda=2\pi/q$ is the light wavelength.

\begin{figure}[t]
 \includegraphics[width=\linewidth]{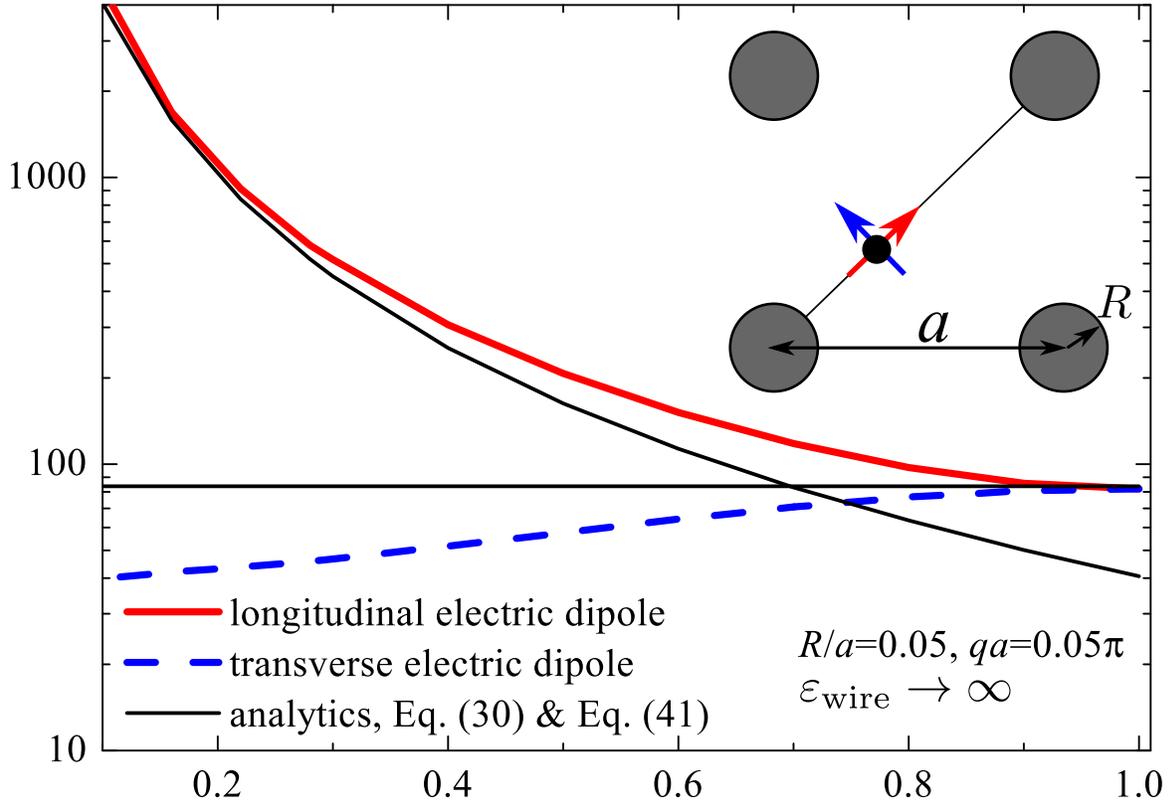}
\caption{(Color online) Purcell factor for electric dipole as function of dipole position within the unit cell. Thick solid and dashed curves are numerical results for longitudinal and transverse dipole orientation (see geometry scheme in the inset). Thin lines correspond to  analytical results Eq.~\eqref{eq:f_e_pos} and  Eq.~\eqref{eq:f_e_center}.  Calculation was performed for $\varepsilon_{\rm wire}\to \infty$, $qa=0.05\pi$ and $R/a=0.05$.}\label{fig:move}
\end{figure}

Comparison between analytical results and numerical calculation is presented in Fig.~\ref{fig:move}. Figure demonstrates high sensitivity of the Purcell factor to the position and orientation of the dipole.
Purcell factor for electric dipole, oriented along the radius vector, greatly increases when the dipole approaches to the center of the wire. This growth is well described by Eq.~\eqref{eq:f_e_pos}, see thin curve in Fig.~\ref{fig:move}. In case of the transverse orientation  the Purcell factor is suppressed for small distances. Thin horizontal line in Fig.~\ref{fig:move} presents analytical answer \eqref{eq:f_e_gen_x} for Purcell factor of the electric dipole in the lattice center, which is obtained in the following section and perfectly agrees with numerical calculation.
\subsection{Effect of wire dielectric constant}\label{sec:epsilon}	
Fig.~\ref{fig:move} shows that large Purcell factor may be obtained even for the dipole in the center of the unit cell. Here we focus on this case, $x_0=y_0=a/2$ and analyze the dependence of the Purcell factor on the dielectric constant of the wires.
In this case to obtain the Green function it is sufficient to keep only the terms with
 $\bm b=0$ in the spectral representation of Eq.~\eqref{eq:Gk}, which reads
\begin{equation}\label{eq:Gek2}
 \bm G^{(e)}_{\bm k}(\boldsymbol \rho)=-\frac{4\pi}{a^2}\sum\limits_{\bm b}\frac{(q^2-k_z^2)\hat{\bm z} -k_z(\bm k_\perp+\bm b)}{q^2-k_z^2-(\bm k_\perp+\bm b)^2}\e^{\rmi(\bm k_\perp+\bm b)\boldsymbol \rho}\:,
\end{equation}
\begin{equation}\label{eq:Gmk2}
 \bm G^{(m)}_{\bm k}(\boldsymbol \rho)=-\frac{4\pi q}{a^2}\sum\limits_{\bm b}\frac{\hat{\bm z}\times(\bm k_\perp+\bm b)}{q^2-k_z^2-(\bm k_\perp+\bm b)^2}\e^{\rmi(\bm k_\perp+\bm b)\boldsymbol \rho}\:.
\end{equation}
 Relatively simple analytical expression for the Purcell factor can be obtained if only the residue at the wavevector of quasi-TEM mode with $k_z\equiv k_{1,z}$ is taken into account in Eq.~\eqref{eq:f}.
 Using Eq.~\eqref{eq:nice3} we perform integration over $k_z$ as follows:
\begin{equation}\label{eq:int}
 \int\frac{dk_z}{2\pi}\frac{F(k_z)}{
(1/\alpha)-C-\rmi 0
}=\rmi\frac{a^2q_p^2}{4\pi}
F(k_{1,z})
\frac{k_{\rm TM}^2+q_p^2-k_{1}^2}{2k_{1}(k_{2}^2-k_{1}^2)}\:.
\end{equation}
As a result,  the Purcell factor is reduced to the integrals over
dimensionless in-plane wavevector $x=k_{\perp}/q_p$:
\begin{eqnarray}\label{eq:fxyz}
 f^{(e)}_{x}=&f^{(e)}_{y}=\frac{3q_p^2}{8q^2}
\Re\!\!\int\limits_0^{(K_{\rm max}/q_p)^2}\rmd x\frac{\sqrt{1+\nu x}}{(1+x)^{3/2}},\\\nonumber
&f^{(e)}_{z}=\frac{3 \varkappa^4}{8q^4}\Re\!\!\int\limits_0^{(K_{\rm max}/q_p)^2}\frac{x\rmd x}{(1+x)^{5/2}\sqrt{1+\nu x}}\:,\\\nonumber
f^{(m)}_{x}=&f^{(m)}_{y}=\frac{3q_p^2}{8q^2}\Re\!\!\int\limits_0^{(K_{\rm max}/q_p)^2}\frac{\rmd x}{\sqrt{1+x}\sqrt{1+\nu x}}\:,\\
&f^{(m)}_z=1\nonumber\:.
\end{eqnarray}
Here the coefficient
\begin{equation}
 \nu=1+\frac{\varkappa^2}{q^2}\equiv 1+\frac{q_p^2}{q^2}\frac{a^2}{\pi(1-\eps_{\rm wire})R^2}
\end{equation}
accounts for the finite dielectric constant of the wires.
\begin{figure}[t]
 \includegraphics[width=\linewidth]{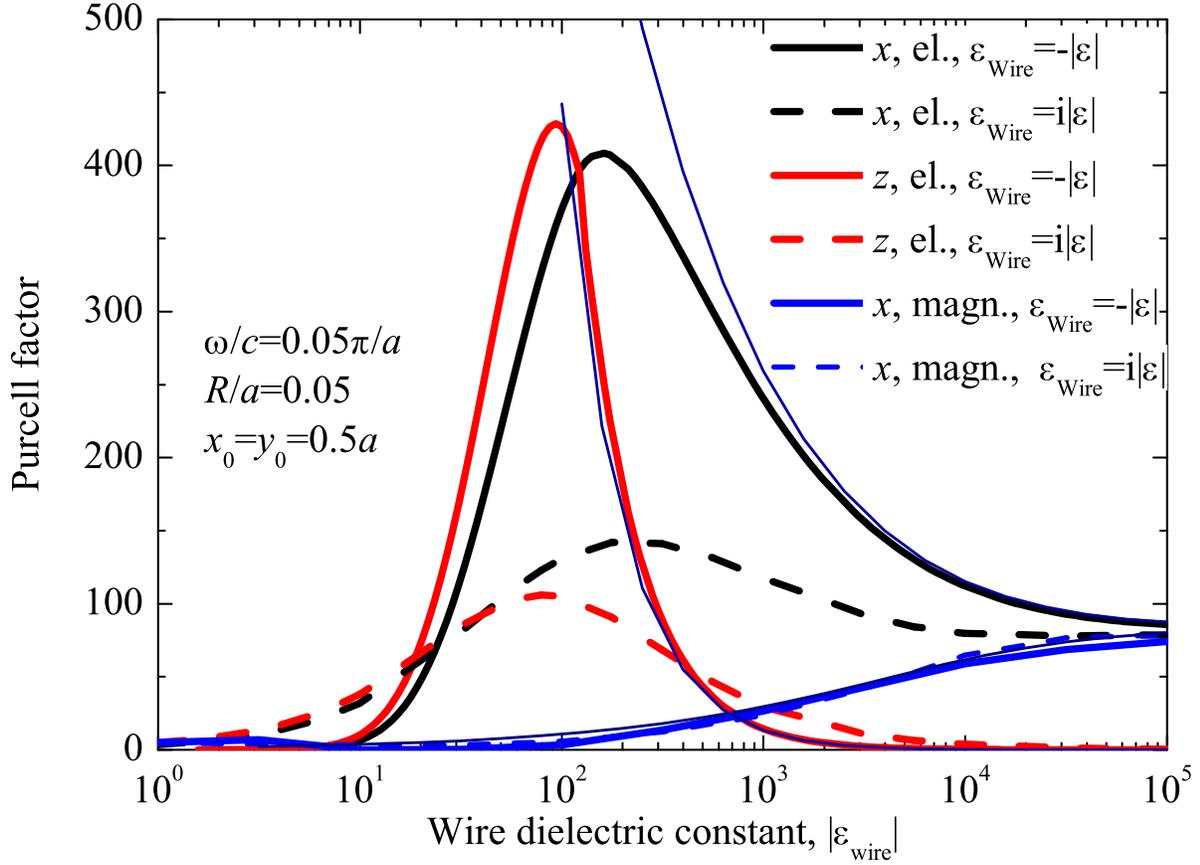}
\caption{(Color online) Purcell factor for electric and magnetic dipoles as function of the wire dielectric constant.
Thick black, red and blue lines correspond to in-plane electric, vertical electric and in-plane magnetic dipole, respectively. Solid and dashed lines are calculated for $\varepsilon_{\rm wire}=\rmi\eps_{\rm wire}$  and $\eps_{\rm wire}=-|\eps_{\rm wire}|$. Thin lines present analytical results \eqref{eq:f_e_gen_x}--\eqref{eq:f_m_gen_x}.
  Calculation was performed at $qa=0.05\pi$, $R/a=0.05$, and $x_0=y_0=a/2$.}\label{fig:plasma}
\end{figure}
The answers after the integration read
\begin{eqnarray}\label{eq:f_e_gen_x}
f_{x,y}^{(e)}&=\frac{3q_p^2}{8q^2}\Re\left[
-2\frac{\beta}{\alpha}+2+2\sqrt\nu\ln\frac{\alpha\sqrt{\nu}+\beta}{\sqrt\nu+1}
\right]\:,\\
f_{z}^{(e)}&=\frac{1}{4}\Re\frac{\beta(\beta^2-3\alpha^2)}{\alpha^3}+\frac{1}{2}\label{eq:f_e_gen_z}\:,\\
 f_{x,y}^{(m)}&=\frac{3q_p^2\label{eq:f_m_gen_x}}{8q^2}\Re\frac2{\sqrt{\nu}}\left[\ln(\sqrt{\nu}\alpha+\beta)-\ln(\sqrt{\nu}+1)\right]\:.
\end{eqnarray}
In the limit of perfect wires
$\eps_{\rm wire}\to\infty$ they reduce to
\begin{eqnarray}
f_{x,y}^{(e)}&=f_{x,y}^{(m)}=\frac{3}{8}\frac{q_p^2}{q^2}\ln\left(1+\frac{K_{\rm max}^2}{q_p^2}\right)\label{eq:f_e_center}\:,\\
f_{z}^{(e)}&=0\:,\\
f_{z}^{(m)}&=1\:.
\end{eqnarray}
Since $q_p\sim 1/a$, the Purcell factor can be estimated as $(\lambda/a)^2$. There is no local field enhancement effect for the dipole located in the cell center, cf. Eq.~\eqref{eq:f_e_pos} and Eq.~\eqref{eq:f_e_center}.

Calculated dependence of electric and magnetic Purcell factor on the wire dielectric constant is presented in Fig.~\ref{fig:epsilon}. The figure demonstrates, that the Purcell factor for in-plane electric dipole unexpectedly grows for finite dielectric constant (black curves). Moreover, spontaneous emission for the vertical electric dipole becomes possible and its rate strongly increases for smaller values of $\varepsilon_{\rm wire}$. This effect is realized both for superconducting wires ($\Re\varepsilon_{\rm wire}<0$, $\Im \eps_{\rm wire}=0$) and for conducting wires ($\Re \varepsilon_{\rm wire}=0$, $\Im\eps_{\rm wire}>0$), cf. solid and dashed curves in Fig.~\ref{fig:epsilon}.
For relatively large values of $|\varepsilon_{\rm wire}|$ the curves are well described by Eqs.~\eqref{eq:f_e_gen_x},\eqref{eq:f_e_gen_z}.
The origin of this Purcell factor enhancement is explained by a competition of two effects: (i) dependence of the wire electric field $\bm G^{(e)}_{\bm k}$ on the quasi-TEM  mode wavevector $k_z$ and (ii) density of states dependence on $k_z$.
 As is demonstrated by Fig.~\ref{fig:kz} and Fig.~\ref{fig:plasma}a, the values of $k_z$ for quasi-TEM modes become larger for non-perfect wires.
Eq.~\eqref{eq:Gek} shows, that both in-plane and axial components of the Green function increase with $k_z$. The density of states for given value $k_{\perp}$ effectively decreases with $k_z$, which is described by the $1/k_z$ factor in Eq.~\eqref{eq:int}. Since the Purcell factor is proportional to the square of the Green function times density of states, is still grows for smaller value of $\varepsilon_{\rm wire}$. The optimum wire of Purcell factor is reached at the dielectric constant $|\eps_{\rm wire}|\sim a^2/R^2$, corresponding to the condition of vanishing effective plasma frequency $q_p(\eps_{\rm wire})$, cf. Fig.~\ref{fig:plasma} and Fig.~\ref{fig:epsilon}. The situation is different for magnetic dipole. Contrary to the electric dipole case, the in-plane components of the corresponding Green function lack the pre-factor $k_z$, see Eq.~\eqref{eq:Gmk}. Consequently, the Purcell factor for in-plane magnetic  dipole  is quenched for smaller values of wire dielectric constant, in agreement with Fig.~\ref{fig:epsilon}(blue curves).
\section{Application to particular structures}\label{sec:Zayats}
\begin{figure}[t]
 \centering\includegraphics[width=0.7\textwidth]{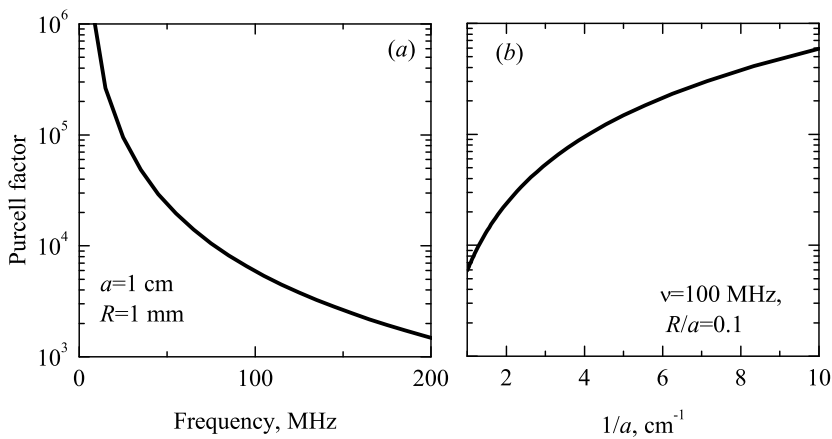}
 \caption{Purcell factor for  wire metamaterial, corresponding to Ref.~\cite{belov2002b} and operating in microwave frequency range. (a) Frequency dependence for $a=1$~cm, $R=0.1a$. (b) Dependence on lattice constant $a$ for
$R=0.1a$ and $\nu=0.1$~GHz.}\label{fig:Micro}
\end{figure}

In this section we apply our general theory to realistic experimental structures, operating in microwave\cite{belov2008trans} and optical\cite{atkinson2006} frequency ranges.
The structure from Ref.~\cite{belov2008trans}, used for subwalength transmission of images, is characterized with the period
$a=1~\rm cm$ and wire radius $R=1~\rm mm$. Asymptotic answer
 Eq.~\eqref{eq:f_e_center} for perfectly conducting wires is valid for microwaves. This answer may be explicitly written as
\begin{equation}
 f_{x,y}^{e}=f_{x,y}^{m}=
\frac{3\pi c^2}{4\omega^2a^2}\frac{\ln[1+\pi/3+2\ln(a/2\pi R)]}{\ln(a/2\pi R)+\pi/6}\:.
\label{eq:micro}
\end{equation}
First factor in Eq.~\eqref{eq:micro}  mainly determines the dependence of the Purcell factor on structure parameters, while the second factor weakly depends on the relative thickness of the wires $R/a$.
Dependence of the Purcell factor on the the source frequency and on the lattice constant is shown in Fig.~\ref{fig:Micro}(a) and  Fig.~\ref{fig:Micro}(b), respectively. Purcell factor strongly increases for smaller frequencies and smaller lattice periods.
Calculation demonstrates that the wire medium allows to achieve high values of Purcell factor in the whole microwave spectral range.
\begin{figure}[t]
\centering\includegraphics[width=0.7\textwidth]{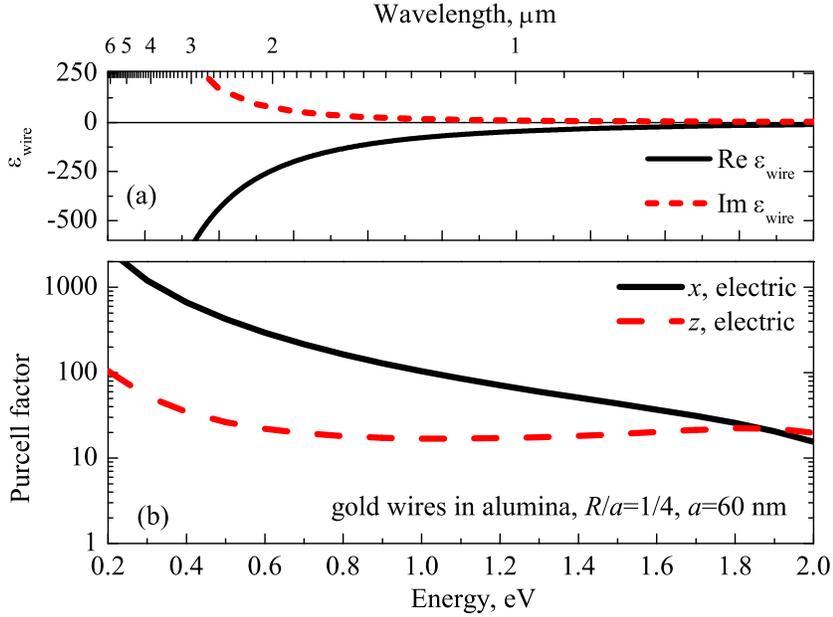}
 \caption{Purcell factor for  nanowire metamaterial, corresponding to Ref.~\cite{atkinson2006}. (a) Energy dependence of the dielectric constant of the wires. (b) Energy dependence of the Purcell factor for horizontal (black solid curve) and vertical (red dashed curve) electric dipole. Calculated for $a=60~$nm, $R=a/4$, $x_0=y_0=a/2$. Decay rate is normalized to the value in bulk matrix.}\label{fig:Zayats}
\end{figure}

In the optical frequency range wire metamaterial may be realized as an array of gold nanowires in alumina~\cite{atkinson2006,Wurtz2008}. Purcell factor calculated for this structure is presented in Fig.~\ref{fig:Zayats}. Realistic energy dependence of the gold dielectric constant of nanowires from Ref.~\cite{atkinson2006} has been taken into account. We have used the value $\varepsilon_{\rm out}=2.56$ for the dielectric constant of the alumina matrix. To account for the dielectric constant of the matrix, different from unity, wire dielectric constant in Eq.~\eqref{eq:epsilon} was scaled as
$\varepsilon_{\rm wire}\to\varepsilon_{\rm wire}/\varepsilon_{\rm out}$, which yields  the spontaneous decay enhancement with respect to the bulk matrix.
Fig.~\ref{fig:Zayats} shows, that for axial dipole orientation a flat maximum in the Purcell factor with $f\sim 20$ is reached at the energy $E\sim 1.8~\rm eV$. This roughly corresponds to the optimal dielectric constant of the wires, revealed in Fig.~\ref{fig:epsilon}. For in-plane
dipole the Purcell factor monotonously decays with photon energy due to decreasing  photonic density of states $\sim 1/(qa)^2$, although its value stays above $10$ for all energies below 2 eV. Several issues should be noted regarding the relevance of Fig.~\ref{fig:Zayats} to real experimental structure. First, our theory has been derived for thin wires, neglecting their transverse polarization. This assumption certainly fails at large energies.
Second,  the real nanorods have the lengths in order of hundreds of nm, while in theory they are assumed infinite. Third, the Purcell factor calculated according to Eq.~\eqref{eq:PurcEM} corresponds to the total decay rate caused by electromagnetic interaction with the medium. It is determined by a sum of the rate of photon radiation in the far field and the rate, with which photons are radiated and then reabsorbed due to the dielectric losses~\cite{Glazov2011}. Thus, Eq.~\eqref{eq:PurcEM} overestimates the enhancement of radiation efficiency. Consequently, curves in Fig.~\ref{fig:Zayats} should be considered as an upper boundary of the Purcell factor in gold nanorod metamaterial, rather than as a rigorous modeling. Still, they indicate that relatively high Purcell
 factor may be achieved in the visible spectral range.


\section{Conclusions}\label{sec:concl}

We have developed a general analytical theory of spontaneous emission of both electric and magnetic dipole sources in wire metamaterials. Our theory goes beyond the effective medium approximation and fully accounts for the discreteness of the structure.  We have analyzed the dependence of the Purcell factor on the dipole position within the lattice unit cell as well as on wire dielectric constant. We have demonstrated that the Purcell factor can be greatly enhanced due to the large density of states of TEM modes, and its values are of the order of a square of the ratio of the light wavelength  and lattice constant. The Purcell factor is also very sensitive to the position and orientation of the dipole source, and it may increase due to the local field effect when the dipole approaches the wires. Counterintuitively, the spontaneous emission rate for the electric dipole emission grows when the wires are not perfectly conducting. We have found optimal value of the wire dielectric constant that maximizes the Purcell factor. We have demonstrated the possibility of broadband spontaneous decay rate enhancement in realistic wire metamaterials operating in both microwave and optical spectral range.

\ack

This work has been supported by the Ministry of Education and Science of Russian Federation, the ``Dynasty'' Foundation, Russian Foundation for Basic Research, European project POLAPHEN, EPSRC~(UK),  and the Australian Research Council. The authors acknowledge useful discussions with C.R. Simovski.

\bibliographystyle{iopart-num}
\providecommand{\newblock}{}


\end{document}